\documentclass[aps,prl,twocolumn,showpacs,superscriptaddress]{revtex4}  
\usepackage{graphicx}  
\usepackage{dcolumn}   
\usepackage{bm}        
\usepackage{amssymb}   

\newcommand{\om}{\ifmmode {w} \else {$w$}\fi}

\newcommand{\BABARPubYear}    {09}
\newcommand{\BABARPubNumber}  {009}

\newcommand{\SLACPubNumber} {13580}

\input babarsym

\begin{document}

\begin{flushleft}
\babar-PUB-\BABARPubYear/\BABARPubNumber \\
SLAC-PUB-\SLACPubNumber \\
\end{flushleft}

\title{Measurement of {\boldmath $|V_{cb}|$} and the Form-Factor Slope in {\boldmath $\Bbar \to D\ell^-\bar{\nu}_{\ell}$} Decays in  Events Tagged by a Fully Reconstructed {\boldmath $B$} Meson}

%
\author{B.~Aubert}
\author{Y.~Karyotakis}
\author{J.~P.~Lees}
\author{V.~Poireau}
\author{E.~Prencipe}
\author{X.~Prudent}
\author{V.~Tisserand}
\affiliation{Laboratoire d'Annecy-le-Vieux de Physique des Particules (LAPP), Universit\'e de Savoie, CNRS/IN2P3,  F-74941 Annecy-Le-Vieux, France}
\author{J.~Garra~Tico}
\author{E.~Grauges}
\affiliation{Universitat de Barcelona, Facultat de Fisica, Departament ECM, E-08028 Barcelona, Spain }
\author{M.~Martinelli$^{ab}$}
\author{A.~Palano$^{ab}$ }
\author{M.~Pappagallo$^{ab}$ }
\affiliation{INFN Sezione di Bari$^{a}$; Dipartimento di Fisica, Universit\`a di Bari$^{b}$, I-70126 Bari, Italy }
% ...
\collaboration{The \babar\ Collaboration}
\noaffiliation

\date{\today}

\begin{abstract}
We present a measurement of the Cabibbo-Kobayashi-Maskawa matrix element $|V_{cb}|$ and the form-factor slope $\rho^2$ in $\Bbar \to D \ell^- \bar{\nu}_{\ell}$ decays based on 
460 million \BB\ events recorded at the $\Upsilon(4S)$ resonance with the \babar\
detector. $\Bbar \to D\ell^-\bar{\nu}_{\ell}$ decays are selected in events in which a hadronic decay of the second $B$ meson is fully reconstructed. We measure the differential decay rate 
and determine ${\cal G}(1) |V_{cb}|= (43.0 \pm 1.9 \pm 1.4)\times 10^{-3}$ and $\rho^2 = 1.20 \pm 0.09 \pm 0.04$, where ${\cal G}(1)$ is the 
 hadronic form factor at the point of zero recoil. 
 We also determine the exclusive branching fractions and find ${\cal B}(\B^- \to D^0 \ell^- \bar{\nu}_{\ell}) = (2.31 \pm 0.08 \pm 0.09)\%$ and ${\cal B} (\Bzb \to D^+ \ell^- \bar{\nu}_{\ell})=(2.23 \pm 0.11 \pm 0.11)\%$.

\end{abstract}

\maketitle 

In the Standard Model (SM) of electroweak interactions, the rate of the semileptonic $\Bbar \to D \ell^- \bar{\nu}_{\ell}$ decay is proportional to the square
of the Cabibbo-Kobayashi-Maskawa (CKM)~\cite{CKM} matrix element $|V_{cb}|$, which is a measure of the weak interaction coupling of the $b$ to the $c$ quark. The length of the side 
of the unitary triangle opposite to the well-measured angle $\beta$ is proportional to the ratio $|V_{ub}/V_{cb}|$, making 
the determination of $|V_{cb}|$ important to test the SM description of \CP\ symmetry violation. 
In addition, imprecise knowledge of $|V_{cb}|$ is an important uncertainty limiting comparison of measurements 
of \CP\ violation in $K$ meson decays with those in $B$ meson decays~\cite{KB}.

Measurements of  $|V_{cb}|$ have been performed in inclusive semileptonic $B$ decays~\cite{incl} and in the exclusive transitions $\Bbar \to D\ell^-\bar{\nu}_{\ell}$ and $\Bbar \to D^*\ell^-\bar{\nu}_{\ell}$~\cite{BaBarBob}.
The most recent inclusive and exclusive determinations differ by more than two standard deviations, with the inclusive result  
more than twice as precise as the exclusive one ~\cite{pdg}. 
Thus improvements in the measurements of exclusive decay rates are highly desirable, particularly for $\Bbar \to D\ell^-\bar{\nu}_{\ell}$ decays, where the experimental uncertainties dominate.
Measurements of $|V_{cb}|$  based on
studies of $\Bbar\to D\ell^-\bar{\nu}_{\ell}$ decays  have previously been reported by the Belle~\cite{BELLE_d}, CLEO~\cite{CLEO_d}, ALEPH~\cite{ALEPH_d} and \babar\ \cite{BaBarBob} Collaborations.  

The  $\Bbar\to D\ell^-\bar{\nu}_{\ell}$ decay rate~\cite{charge} is proportional to the square of the hadronic matrix element 
that describes the effects of strong interactions in $\Bbar \to D$ transitions. In the limit of very small lepton masses ($\ell=e$ or $\mu$), their effect can be parameterized by a single form factor ${\cal G}(\om)$:

\begin{eqnarray}
\label{eq:diffrate_dlnu}                                         
&& \frac{{\rm d}\Gamma(\Bbar \to D\ell\nu)}{{\rm d}\om}  \\
&& = \frac{G^2_F}{48 \pi^3 \hbar} M^3_{D} (M_{B}+M_{D})^2 
                            ({\om^2-1})^{3/2}  \mid V_{cb} \mid^2 ~ {\cal G}^2 (\om),      \nonumber \\ \nonumber
\end{eqnarray}

\noindent where $G_F$ is the Fermi coupling constant, and $M_{B}$ and $M_{D}$ are the masses of the $B$ 
and $D$ mesons, respectively. The variable $\om$ denotes the product of the $B$ and $D$ meson four-velocities $V_B$ and $V_D$, 
$\om = V_B\cdot V_D=(M_{B}^2 + M_{D}^2 - q^2)/(2M_{B} M_{D})$, where $q^2 \equiv (p_{B}-p_{D})^2$, and $p_B$ and $p_D$ are the four-momenta of the $B$ and $D$ mesons.  

In the limit of infinite quark masses, ${\cal G}(\om)$ coincides with the 
Isgur-Wise function~\cite{IW}. This function is normalized to unity at zero recoil, where $q^2$ is maximum. Corrections to the heavy quark limit have been calculated based on unquenched~\cite{Okamoto} and quenched lattice QCD~\cite{Nazario}.
Thus $|V_{cb}|$ can be extracted by extrapolating the differential
decay rate to $\om = 1$. To reduce the uncertainties associated with this extrapolation, constraints on the shape of the form factor are necessary. Several functional forms have been proposed~\cite{Neuold}. 
We adopt the parameterization suggested in Ref.~\cite{Neunew}, where the non-linear dependence of the form factor on $\om$ is expressed 
in terms of a single shape parameter, the form-factor slope $\rho^2$.

In this letter, we present a measurement of the differential decay rates for $\Bzb \to D^+ \ell^- \bar
{\nu}_{\ell}$ and $B^- \to D^0 \ell^- \bar{\nu}_{\ell}$ decays. 
The analysis is based on data collected with the \babar\ detector~\cite{detector} at the 
\pep2\ asymmetric-energy $e^+e^-$ storage rings. The data consist 
of 417~fb$^{-1}$ recorded at the $\Upsilon(4S)$  resonance, corresponding to approximately 460 million \BB\ pairs. An additional sample of  40~fb$^{-1}$, collected at a center-of-mass (CM) energy 40 MeV below the $\Upsilon(4S)$ resonance, is used to study background from $e^+e^- \to f\bar{f}~(f=u,d,s,c,\tau)$ continuum events. We also use samples of {\sc GEANT4} Monte Carlo (MC) simulated events that correspond to about three times the data sample size.
The simulation models $\Bbar \to D^{(*)} \ell^- \bar{\nu}_{\ell}$
decays using calculations based on Heavy Quark Effective Theory (HQET)~\cite{Neunew}, $\Bbar \to D^{**}(\rightarrow D^{(*)} \pi)
 \ell^- \bar{\nu}_{\ell}$ decays using the ISGW2 model~\cite{ISGW}, and non-resonant $\Bbar\to D^{(*)} \pi \ell^- \bar{\nu}_{\ell}$ decays using the Goity-Roberts model~\cite{Goity}.
The MC simulation includes radiative effects such as bremsstrahlung in the detector material. 
QED final-state radiation is modeled by PHOTOS~\cite{photos}.

Semileptonic decays are selected in \BB\ events in which a hadronic decay of the second $B$ meson ($B_\mathrm{tag}$) is fully reconstructed. This leads to a very clean sample of events and also provides a precise measurement of $q^2$ and hence $\om$.  
We first reconstruct the semileptonic $B$ decay, selecting a lepton with momentum in the CM frame $p^*_{\ell}$ larger than 0.6 GeV. For electrons,  we search for a vertex formed in conjunction with a track of opposite charge and remove those with an invariant mass consistent  with a photon conversion or a $\pi^0$ Dalitz decay.  Candidate $D^0$ mesons that have the correct charge correlation with the lepton are reconstructed
in the $K^-\pi^+$, $K^- \pi^+ \pi^0$, $K^- \pi^+ \pi^+ \pi^-$,
$K^0_S \pi^+ \pi^-$, $K^0_S \pi^+ \pi^- \pi^0$, $K^0_S \pi^0$, $K^+ K^-$,
$\pi^+ \pi^-$, and $K^0_S K^0_S$ channels, and $D^+$ mesons in the
$K^- \pi^+ \pi^+$, $K^- \pi^+ \pi^+ \pi^0$, $K^0_S \pi^+$, $K^0_S \pi^+ \pi^0$,
$K^+ K^- \pi^+$, $K^0_S K^+$, and $K^0_S \pi^+ \pi^+ \pi^-$ channels.
$D^0 (D^+)$ candidates are required to be within 2$\sigma$ of the nominal $D^0 (D^+)$ mass, where $\sigma$ is approximately 8 MeV$/c^{2}$. In events with multiple $D\ell^-$ combinations, the candidate with the largest $D\ell^-$ vertex probability is selected. Events with more than one reconstructed lepton with $p^*_{\ell}>0.6$ GeV are vetoed.

 We reconstruct $B_\mathrm{tag}$ decays~\cite{BrecoVub} in charmed hadronic modes $\Bbar \rightarrow D^{(*)} Y$, where 
$Y$ represents a collection of hadrons, composed
of $n_1\pi^{\pm}+n_2 K^{\pm}+n_3 K^0_S+n_4\pi^0$, where $n_1+n_2 =1,3,5$, $n_3
\leq 2$, and $n_4 \leq 2$. Using $D^0(D^+)$ and $D^{*0}(D^{*+})$ as seeds for $B^-(\Bzb)$ decays, we reconstruct about 1000 different decay  modes.
The kinematic consistency of a $B_\mathrm{tag}$ candidate with a $B$ meson decay is evaluated using two variables: the beam-energy
substituted mass $m_{ES} \equiv \sqrt{s/4-(p^*_B)^2}$, and the energy difference $\Delta E \equiv E^*_B -\sqrt{s}/2$. Here $\sqrt{s}$ is the total CM  energy, and $p^*_B$ and $E^*_B$ denote the momentum and energy of the $B_\mathrm{tag}$ candidate in the CM frame. For correctly identified $B_\mathrm{tag}$ decays, $m_{ES}$ peaks at the $B$-meson mass, while $\Delta E$ is consistent
with zero. We select $B_\mathrm{tag}$ candidates in the signal region
defined as 5.27~GeV $< m_{ES} <$ 5.29~GeV, excluding those that have
 daughters in common with the $\Bbar \to D\ell^-\bar{\nu}_{\ell}$ decay. In the case of multiple $B_\mathrm{tag}$ candidates in an event, we select the one with the smallest
$|\Delta E|$ value. The $B_\mathrm{tag}$ and the $D\ell^-$ candidates are required to have the correct charge-flavor correlation. 
Cross-feed events, $i.e.$, $B^-_\mathrm{tag} (\Bzb_\mathrm{tag})$ candidates erroneously reconstructed as a neutral~(charged) $B$ meson,  are subtracted using estimates from the simulation.

Semileptonic $B$ decays are identified by their missing-mass squared value, $m^2_\mathrm{miss} = \left[ p(\Upsilon(4S)) -p(B_\mathrm{tag}) - p(D) - p(\ell)\right]^2$, defined in terms of the measured particle four-momenta.
For correctly reconstructed signal events, the only missing particle is the neutrino and $m^2_\mathrm{miss}$ peaks at zero. Other semileptonic $B$ decays, like  $\Bbar \to D^{(*,**)} \ell^- \bar{\nu}_{\ell}$, where at least one particle is not reconstructed (feed-down), yield larger values of $m^2_\mathrm{miss}$.

We measure $|V_{cb}|$ and the form-factor slope $\rho^2$ by a fit to the $\om$ distribution. We examine the data and MC events in ten equal-size $\om$ bins in the interval $1 < \om < 1.6$.  
  Since the $B$ momentum is known from the fully reconstructed $B_\mathrm{tag}$ in
 the same event, $\om$ can be reconstructed with good precision, namely to $\sim 0.01$, which corresponds to about $2\%$ of the full kinematic range.

To obtain the $\Bbar \to D\ell^-\bar{\nu}_{\ell}$ signal yield in each bin of $\om$, 
we perform a one-dimensional extended binned maximum likelihood fit~\cite{Barlow} to the corresponding $m^2_\mathrm{miss}$ distribution. The fitted data samples are assumed to contain four different types of events: $\Bbar \to D\ell^- \bar{\nu}_{\ell}$ signal events, feed-down from other semileptonic $B$ decays, combinatorial \BB\ and continuum background, and fake lepton events (mostly from hadronic $B$ decays with hadrons misidentified as leptons).
The Probability Density Functions (PDFs) are derived from the MC predictions for the different semileptonic $B$ decay $m^2_\mathrm{miss}$ distributions. We use the off-peak data to provide the continuum background normalization. The shape of the continuum background distribution predicted by the MC simulation is consistent with that obtained from the off-peak data. 

The measured $m^2_\mathrm{miss}$ distributions are compared with the results of the fits for two different $\om$ intervals in Fig.~1.

\begin{figure}[!ht]
\includegraphics[width=0.45\textwidth,totalheight=0.35\textheight]{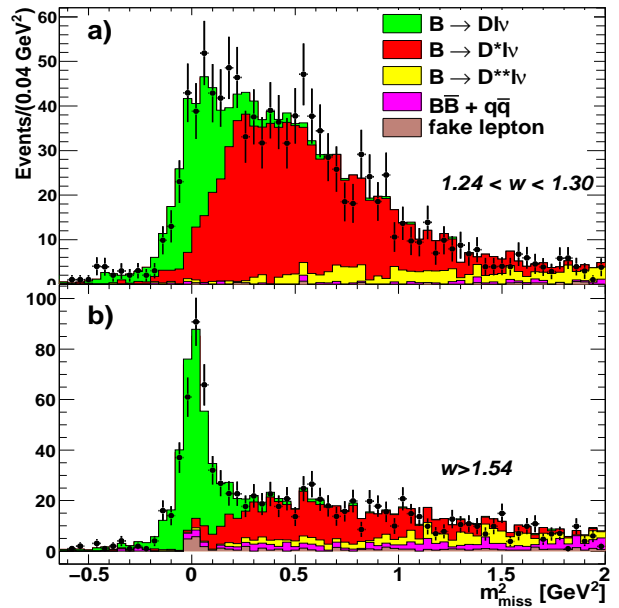}
\caption{(Color online)  Fit to the $m^2_\mathrm{miss}$ distribution, in two different $\om$ intervals, 
for $B^- \to D^0 \ell^- \bar{\nu}_{\ell}$: the data (points with error bars) are compared to the results of the overall fit (sum of the solid histograms). The PDFs for the different fit components are stacked in the order shown in the legend.}
\label{fig:Fit}
\end{figure}

We perform a least-squares fit to the observed signal yields in the ten bins of $\om$. 
We minimize a $\chi^2$ defined as

\begin{equation}
\label{eq:chi2}
\chi^2 = \sum_{i=1}^{10} \frac{(N^i_{\rm data} - \sum_{j=1}^{N^i_{\rm MC}} W^i_j )^2}{ (\sigma^i_{\rm data})^2 +  \sum_{j=1}
^{N^i_{\rm MC}} W{^i_j}^2},
\end{equation}

\noindent 

\noindent where the index $i$ refers to the $\om$ bin and the index $j$ runs over all MC events in bin $i$;~$N^i_{\rm data}$ is the observed number of signal events found in the $i^{th}$ bin and $\sigma^i_{\rm data}$ the corresponding uncertainty. 
The expected signal yields are calculated at each step of the minimization from the reweighted sum of $N^i_{\rm MC}$ simulated events. Each weight is the product of two terms,
 $W^i_j = W^{\cal L} \times W^{i,\rm theo}_j$ where $W^{\cal L}$ is an overall fixed scale factor, which accounts for the relative integrated luminosity of the  
data and signal MC events, and $ W_j^{i,\rm theo} $ is computed using the true $w$ value of the event $j$ and depends on ${\cal G}(1) |V_{cb}|$ and $\rho^2$, which are free parameters determined in the fit that are recalculated at each step of the minimization.

\begin{figure}[!t]
\begin{tabular}{ c c }
\includegraphics[width=0.45\textwidth,totalheight=0.35\textheight]{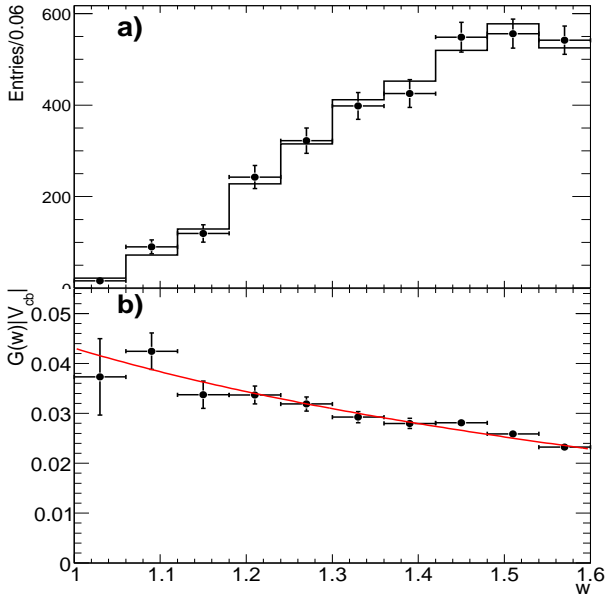}\\
\end{tabular}
\caption{(a) $\om$ distribution obtained summing
 $B^- \to D^0\ell^-\bar{\nu}_{\ell}$ and $\Bzb \to D^+\ell^-\bar{\nu}_{\ell}$ yields. The data (points with error bars) 
are compared to the results of the overall
 fit (solid histogram). (b) ${\cal G}(w)|V_{cb}|$ distribution corrected for the 
reconstruction efficiency, with the fit result superimposed. }
\label{fig:dndw-all} 
\end{figure}

We first fit the $\om$ distributions for the charged and neutral $\Bbar\to D\ell^-\bar{\nu}_{\ell}$ 
samples separately, and then perform a fit to the combined $\Bbar\to D\ell^-\bar{\nu}_{\ell}$ sample. 
In Fig. \ref{fig:dndw-all} we show the comparison between the data and the fit results for the combined sample.
The measured values of ${\cal G}(1)|V_{cb}|$ and $\rho^2$, with the corresponding correlation $\rho_\mathrm{corr}$ obtained 
from the fit,  are reported in Table \ref{t:datafit}. The value of the branching fraction is computed by integrating the 
differential expression in Eq.~\ref{eq:diffrate_dlnu} and dividing by the appropriate $B$-meson lifetime.

\begin{table*}[!]
\centering
\caption{\label{t:datafit} Fit results for each sample. In the last column we report the results for the 
$\Bzb$ and $B^-$ combined fit, where the branching fraction refers to $\Bzb$ decays. We also report the signal yields and the 
reconstruction efficiencies, integrated over the full $w$ range. 
For ${\cal G}(1)|V_{cb}|$ and $\rho^2$, we report both the statistical and systematic uncertainties.}
\begin{tabular}{l|cc|c}
\hline
\hline
                         & $B^- \to D^0\ell^-\bar{\nu}_{\ell}$            & $\Bzb \to D^+\ell^-\bar{\nu}_{\ell}$        
         & $\Bbar \to D\ell^-\bar{\nu}_{\ell}$       \\
\hline 
${\cal G}(1)\Vcb\cdot 10^3$ &41.7$\pm$2.1 $\pm$1.3       &45.6$\pm$ 3.3$\pm$1.6       & 43.0$\pm$ 1.9$\pm$1.4 \\
$\rho^2$                    & 1.14$\pm$  0.11$\pm$0.04      & 1.29$\pm$ 0.14$\pm$0.05     & 1.20$\pm$ 0.09$\pm$0.04\\
$\rho_\mathrm{corr}$               & 0.943                 & 0.950              &   0.952      \\
$\chi^2/ndf$             & 3.4/8                   & 5.6/8                      & 9.9/18                  \\
\hline
Signal Yield & 2147 $\pm$ 69 &  1108 $\pm$ 45 & - \\
Recon. efficiency & $(1.99 \pm 0.02)\times 10^{-4}$  & $(1.09 \pm 0.02) \times 10^{-4}$& -\\
${\cal B}$              &(2.31$\pm$ 0.08 $\pm$ 0.09)$\%$ & (2.23$\pm$ 0.11 $\pm$ 0.11)$\%$    & (2.17$\pm$ 0.06 $\pm$ 0.09)$\%$\\
\hline
\hline
\end{tabular}
\end{table*}

In order to reduce the systematic uncertainty on the measurement of ${\cal G}(1) |V_{cb}|$ and the branching fractions, 
we normalize the exclusive signal yield to the yield of inclusive semileptonic decays, $\Bbar \to X \ell^- \bar{\nu}_{\ell}$, in events tagged by 
a fully reconstructed hadronic $B$ decay. 
The inclusive $\Bbar \to X \ell^- \bar{\nu}_{\ell}$ decays are selected by identifying one charged lepton with $p^*_{\ell}>0.6$ GeV  and the charge expected based on the $B_\mathrm{tag}$ decay. In the case of multiple $B_\mathrm{tag}$ candidates in an event, we select the decay mode with the highest purity, estimated from the MC prediction for the fraction of true decays in the $m_{ES}$ signal region. 
Background components that peak in the $m_{ES}$ signal region include cascade $B$ meson decays, for which the lepton does not come directly from the $B$, and hadronic decays, and are subtracted using the corresponding MC predictions.  
The inclusive $\Bbar \to X \ell^- \bar{\nu}_{\ell}$ yield is obtained from a maximum likelihood fit to the $m_{ES}$ distribution of the $B_\mathrm{tag}$ candidates, as described in Ref.~\cite{babar-3}. The fit yields $(198.9 \pm 1.6) \times 10^3$ events for the 
$B^- \to X \ell^- \bar{\nu}_{\ell}$ sample and $(116.3 \pm 1.0) \times 10^3$ events for the $\Bzb \to X \ell^- \bar{\nu}_{\ell}$ sample.  The corresponding reconstruction efficiencies,
including the $B_\mathrm{tag}$ reconstruction, are 0.39\% and 0.25\%, respectively.

Numerous sources of systematic uncertainty are investigated.
The largest uncertainties are due to differences in the efficiency of the $B_{tag}$ selection between the exclusive $\Bbar \to D \ell^- \bar{\nu}_{\ell}$ and inclusive $\Bbar \to X \ell^- \bar{\nu}_{\ell}$ decays (a relative 1.5\% systematic uncertainty on $|V_{cb}|$),  the $\Bbar \to D \ell^- \bar{\nu}_{\ell}$ fit procedure (1.3\%), and the uncertainties on the branching fractions of the reconstructed $D$ decay  modes and $\Bbar \to D^{**}\ell^-\bar{\nu}_{\ell}$ decays (1.1\%).
The uncertainties due to the detector simulation are established by varying, within bounds given by data control samples, the tracking efficiency of all charged tracks (0.7\%), the calorimeter efficiency (0.9\%), and the lepton identification efficiency (0.9\%).  
We evaluate the systematic uncertainties associated with the MC simulation of various signal and background processes: photon conversion and $\pi^0$ Dalitz decay, $B$ cascade decay contamination (0.8\%), and  flavor cross-feed (0.2\%). 
The uncertainty arising from radiative corrections (0.1\%) is studied by comparing the standard results with those obtained when
 PHOTOS is not used. We take 30\% of the difference as a conservative systematic uncertainty. 
We vary the $\Bbar \to D^{*} \ell^- \bar{\nu}_{\ell}$ form factors (0.4\%) within their measured uncertainties~\cite{BaBarBob} and 
 use an HQET parameterization~\cite{LLSW} to describe $\Bbar \to D^{**} \ell^- \bar{\nu}_{\ell}$ decays (0.3\%).  
We evaluate an uncertainty associated with the $\Bbar \to X \ell^- \bar{\nu}_{\ell}$ fitting procedure (0.8\%), and with the absolute branching fraction ${\cal B} (\Bbar \to X \ell^- \bar{\nu}_{\ell})$ used for the normalization (0.8\%). The complete set of systematic uncertainties is given in Ref.~\cite{epaps}.

From the fit to the combined $\Bbar \to D \ell^- \bar{\nu}_{\ell}$ sample,  we measure ${\cal G}(1)|V_{cb}|=(43.0 \pm 1.9 \pm 1.4)\times 10^{-3}$. Using an unquenched lattice calculation \cite{Okamoto}, corrected by a factor of 1.007 for QED effects, we obtain $|V_{cb}|=(39.8\pm 1.8 \pm 1.3 \pm 0.9_{FF})\times 10^{-3}$,  
where the third error is due to the theoretical uncertainty in ${\cal G}(1)$. As an alternative, we use a quenched lattice calculation based on the Step Scaling Method (SSM)~\cite{Nazario}, and obtain $|V_{cb}|=(41.6 \pm 1.8 \pm 1.4 \pm 0.7_{FF})\times 10^{-3}$.  
The authors of~\cite{Nazario} report the lattice determination of ${\cal G}(\om)$ for finite momentum transfer. Although quenched, this new calculation allows the extraction of $|V_{cb}|$ with relatively small model dependence avoiding the large extrapolation to $\om=1$. For example, from a linear interpolation around $w=1.2$, we obtain $|V_{cb}|=(41.4 \pm 1.3 \pm 1.4 \pm 1.0_{FF})\times 10^{-3}$. We report our measurements of ${\cal G}(\om)|V_{cb}|$ for $\om >1$ in Ref.~\cite{epaps}.  

The resulting value of $|V_{cb}|$ extrapolated to $w=1$ is largely independent of previous \babar\ results~\cite{BaBarBob} and significantly more precise than previous measurements from $\Bbar \to D\ell^-\bar{\nu}_{\ell}$ decays. It is also consistent with the measurement obtained from $\Bbar \to D^*\ell^-\bar{\nu}_{\ell}$ decays and with the inclusive determination
of $|V_{cb}| = (41.6 \pm 0.6) \times 10^{-3}$~\cite{pdg}.

We are grateful for the excellent luminosity and machine conditions
provided by our \pep2\ colleagues, 
and for the substantial dedicated effort from
the computing organizations that support \babar.
The collaborating institutions wish to thank 
SLAC for its support and kind hospitality. 
This work is supported by
DOE
and NSF (USA),
NSERC (Canada),
CEA and
CNRS-IN2P3
(France),
BMBF and DFG
(Germany),
INFN (Italy),
FOM (The Netherlands),
NFR (Norway),
MIST (Russia),
MEC (Spain), and
STFC (United Kingdom). 
Individuals have received support from the
Marie Curie EIF (European Union) and
the A.~P.~Sloan Foundation.

\clearpage
















\begin{table*}
\flushleft
\textbf{\large Electronic Physics Auxiliary Publication Service (EPAPS)}\\
\smallskip
\normalsize
This is an EPAPS attachment to B.~Aubert \textit{et al.} (\babar\ Collaboration),
\babar-PUB-\BABARPubYear/\BABARPubNumber, SLAC-PUB-\SLACPubNumber, 
submitted to Phys.\ Rev.\ Lett. For more information on EPAPS, see
http://www.aip.org/pubservs/epaps.html.
\end{table*}

\begin{table*}[!h]
\centering
\caption{Systematic uncertainties in the measurement of ${\cal G}(1)|V_{cb}|$, $\rho^2$ and the branching
fraction for $\overline{B} \rightarrow D \ell^- \bar{\nu}_{\ell}$ decays. We report the relative error (in \%)  for ${\cal G}(1)|V_{cb}|$ and branching fraction, and the absolute error on $\rho^2$.}
\begin{tabular}{|l|c|c|c|c|c|c|c|c|c|}
\hline
\hline
 &\multicolumn{9}{c|}{{\scriptsize Systematic uncertainty on $|V_{cb}|$, $\rho^2$ and BF}} \\
\hline
& \multicolumn{3}{c|}{$D^0\ell^-\bar{\nu}_{\ell}$} & \multicolumn{3}{c|}{$D^{+}\ell^-\bar{\nu}_{\ell}$} & \multicolumn{3}{c|
}{$D\ell^-\bar{\nu}_{\ell}$} \\
\hline
& $|V_{cb}| (\%)$ & $\rho^2$ & $BF$ (\%) & $|V_{cb}| (\%)$ & $\rho^2$ & $BF$ (\%) & $|V_{cb}| (\%)$ & $\rho^2$ & $BF$ (\%)\\
\hline
Tracking efficiency      & 0.5 &0.008   & 0.7 & 1.1  & 0.003 & 1.4 & 0.7 & 0.004 &1.0\\
Neutral reconstruction   & 1.0 &0.003    & 1.2 & 0.8  & 0.006 & 0.9 & 0.9 & 0.004 &1.2 \\
Lepton ID                & 1.0 & 0.009  & 1.0 & 0.9  & 0.009 & 0.8 & 0.9& 0.009 &0.9 \\
Final State Radiation    & 0.1 & 0.005 & 0.2 & 0.1 & 0.005 & 0.2 & 0.1 &  0.005 &0.2 \\
Cascade $\Bbar \to X \to \ell^-$ decay background  & 0.6 & - & 1.2& 1.0 & - & 2.0 & 0.8 & - &1.5\\
$B^0-B^{\pm}$ cross-feed  & 0.2 & 0.003 & 0.2 & 0.2 & 0.003 & 0.2 & 0.2 & 0.003 & 0.2\\
$\Bbar \to D^{*}\ell^-\bar{\nu}_{\ell}$ form factors &0.6 &0.008 & 0.5 & 0.2 & 0.003 & 0.2 & 0.4  & 0.006 & 0.3 \\
$\Bbar \to D^{**}\ell^-\bar{\nu}_{\ell}$ form factors & 0.2 &0.007& 0.2 & 0.3 & 0.006 & 0.2 & 0.3  &0.007& 0.1 \\
$D$ branching fractions & 1.0 &-& 2.0 & 1.4 & - & 2.7 & 1.1 &- & 2.2\\
${\cal B}(\Bbar \to D^{**}\ell^-\bar{\nu}_{\ell})$ & 1.2 &0.023& 0.6 & 1.0 & 0.011 & 0.9 & 1.1  &0.019 & 0.6 \\
${\cal B} (\overline{B} \to X \ell^- \bar{\nu}_{\ell})$ & 0.9  & - & 1.9 & 0.9 & -  & 1.9 & 0.8 & - &1.7\\
$B_\mathrm{tag}$ selection & 1.1 & 0.021 & 0.6 & 1.8 & 0.036 & 0.8 & 1.5 & 0.028 &0.8\\
$\Bbar \to X \ell^- \bar{\nu}_{\ell}$ fit  &0.7 & - & 1.4 & 1.1 & - & 2.2 & 0.8 & - & 1.7\\
$\Bbar \rightarrow D \ell^- \bar{\nu}_{\ell}$ fit &1.3 & 0.018 & 1.1 & 1.1 & 0.027 & 0.6 & 1.3 & 0.020 & 0.8\\
$B$ meson lifetime     & - & - & 0.7 & - & - & 0.6 & - & - & 0.6\\
\hline
 Total systematic error & 3.1 & 0.04 & 4.1 & 3.6 & 0.05 & 5.0 & 3.3& 0.04 & 4.3 \\
\hline
\hline
\end{tabular}
\label{tab:Syst1}
\end{table*}

\begin{table*}[!h]
\centering
\caption{\label{t:fitw} Fit results for $|V_{cb}|{\cal{G}}(\om)$ 
extracted at different value of $\om$. In order to reduce any fit model
dependence, we fit the data interpolating
few bins around $\om$. In particular we use 4 bins between $\om=1.00$ and $\om=1.24$ 
to extract $|V_{cb}|{\cal{G}}(\om)$ at $\om=1.03,1.05$ and $1.10$, and 4 bins between 
$\om=1.06$ and $\om=1.30$ to extract $|V_{cb}|{\cal{G}}(\om=1.20)$.
The statistical correlation between the measurement at $\om=1.20$ and the others is $\rho_{\rm corr}=0.57$,
instead the systematic uncertainties can be assumed correlated at 100\%.
In the last column we report the results for $|V_{cb}|$ obtained using the results for 
the ${\cal{G}}(\om)$ computed in G.M. de Divitiis {\it et al.}, Phys.~Lett.~B{\bf 655}, 45 (2007). }
\begin{tabular}{l|cc|c}
\hline
\hline
$\om$ & $|V_{cb}|\cdot {\cal{G}}(\om)\cdot 10^{3}$ & ${\cal{G}}(\om)$  & $|V_{cb}|\cdot 10^{3}$   \\
\hline 
 1.03     & 40.9$\pm$5.7$\pm$1.3    & 1.001$\pm$0.019   & 40.9$\pm$5.7$\pm$1.3$\pm$0.8 \\       
 1.05     & 40.2$\pm$5.0$\pm$1.3    & 0.987$\pm$0.015   & 40.7$\pm$5.1$\pm$1.3$\pm$0.6 \\ 
 1.10     & 38.3$\pm$3.3$\pm$1.3    & 0.943$\pm$0.011   & 40.6$\pm$3.5$\pm$1.4$\pm$0.5 \\ 
 1.20     & 35.3$\pm$1.1$\pm$1.2    & 0.853$\pm$0.021   & 41.4$\pm$1.3$\pm$1.4$\pm$1.0 \\
\hline
\hline
\end{tabular}
\end{table*}



\begin{thebibliography}{99}
\bibitem{charge}
The charge conjugate state is always implied unless stated otherwise.
\end{thebibliography}
\end{document}